\documentstyle[amssymb,prl,aps,twocolumn,epsf]{revtex} 
\draft
\newcommand{\be}{\begin{equation}} 
\newcommand{\ee}{\end{equation}} 
\newcommand{\bea}{\begin{eqnarray}} 
\newcommand{\eea}{\end{eqnarray}}

\begin{document} 
\twocolumn[\hsize\textwidth\columnwidth\hsize\csname @twocolumnfalse\endcsname

\title {Dispersion of the neutron resonance in cuprate superconductors}  
\author{Andrey V. Chubukov,$^{1,2}$ Boldizs\'ar Jank\'o,$^{1,3}$ 
and  Oleg Tchernyshyov$^{1,4}$}   
\address{$^1$ Materials Sciences Division, Argonne National
  Laboratory, Argonne, Illinois 60439} 
\address{$^2$Department of Physics, University of Wisconsin, 
Madison, Wisconsin 53706}   
\address{$^3$Physics Department,  University of Notre Dame, 
Notre Dame, Indiana 46556-5670}
\address{$^4$School of Natural Sciences, Institute for Advanced Study, 
Princeton, New Jersey 08540}   

\date{\today}   
\maketitle    
\begin{abstract}   
We argue that recently measured downward dispersion of the neutron
resonance peak in cuprate superconductors is naturally explained if
the resonance is viewed as a spin-1 collective mode in a d-wave
superconductor.  The reduction of the resonant frequency away from the
antiferromagnetic wave vector is a direct consequence of the momentum
dependence of the d-wave superconducting gap.  When the magnetic
correlation length becomes large, the dispersion should become
magnon-like, i.e., curve upwards from $(\pi,\pi)$.
\end{abstract}  

\pacs{PACS numbers: 71.10.Ca, 74.20.Fg, 74.25.-q}   
] 
 
\narrowtext 

Superconductivity and antiferromagnetism are two major
phases of the high-$T_c$ cuprates.  When viewed separately, the
corresponding ground states appear to be quite conventional: the
parent compounds of high $T_c$ cuprates (such as La$_2$CuO$_4$
\cite{Mason}) are exemplary Heisenberg antiferromagnets, while
overdoped cuprates in many respects resemble BCS-type d-wave
superconductors\cite{overd}.  However, full understanding of the
interplay between the two phenomena is sorely lacking.  Although most
experts agree that antiferromagnetism is the ultimate cause of
high-$T_c$ superconductivity, the intermediate steps are not yet
clear~\cite{discussion}.

Perhaps the strongest experimental indication of the interplay
between antiferromagnetism and superconductivity in cuprates is 
the discovery of strong inelastic neutron scattering  deep in the superconducting
(SC) phase of materials with the highest $T_c$.  The most intense 
scattering at $T \ll T_c$ has been detected in YBCO \cite{RM} 
and Bi2212 \cite{Keimer2} near the antiferromagnetic (AF) wave vector  
${\bf q = Q} =(\pi,\pi)$ and energy $\Omega_{\bf Q} \approx 40$ meV.
Experiments conducted with polarized neutrons \cite{Mook1} 
indicate that the resonant scattering is due to electron spins.  

\begin{figure}[tbp]
\centerline{
\epsfxsize=\columnwidth
\epsffile{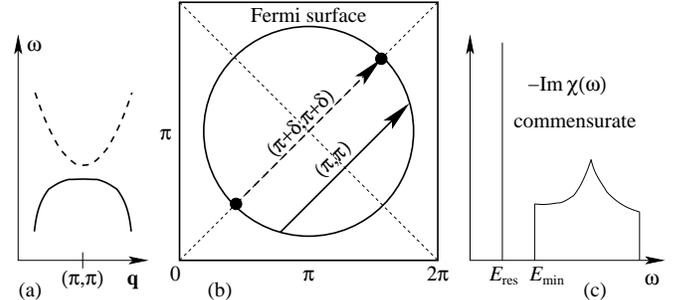}}
\caption{(a) A sketch of the resonance dispersion: experimental (solid
line), a massive magnon in the disordered N\'eel state (dashed line).
(b) Approximate shape of the Fermi surface in Bi2212.  Filled dots are
nodal points, where the fermions are gapless.  The continuum of
two-particle excitations at the corresponding wave vector ${\bf
q}_0=(\pi+\delta,\pi+\delta)$ (dashed arrow) starts at zero frequency.
For lattice momentum ${\bf Q}=(\pi,\pi)$ (solid arrow) the continuum
starts at a finite energy.  (c) spin susceptibility at the wave vector
$(\pi,\pi)$.}
\label{fig0}
\end{figure}

It is tempting to interpret the resonance as a magnon in a disordered
N\'eel state, which becomes more visible in a superconductor. 
 However, a magnon frequency 
\begin{equation}
\Omega_{\bf q}^2 = c^2(\xi^{-2} + |{\bf q-Q}|^2) 
= \Omega_{\bf Q}^2 + c^2|{\bf q-Q}|^2,
\label{eq:massive-magnon}
\end{equation} 
where $c$ is a spin-wave velocity, clearly {\it increases} away from
the AF wave vector ${\bf Q}=(\pi,\pi)$~\cite{subir}.  Meanwhile,
recent experiments on YBCO~\cite{Arai,Keimer3} have revealed the
opposite trend: the resonance energy {\em decreases} away from
$(\pi,\pi)$ [see Fig. \ref{fig0}(a)].  In a disordered
antiferromagnet, this can only be the case if spin response in the
normal state is incommensurate \cite{subir2}.  However, the data
suggest\cite{Keimer3} that, unlike in $214$ materials
\cite{n-st-neutrons}, the normal-state spin response in YBCO is
commensurate.
 
On second thought, the best superconductors of the cuprate family
differ significantly from their parent AF compounds.  They have a
large, Luttinger-like Fermi surface in the normal state, and fermionic
quasiparticles in the SC state, as shown by angle-resolved
photoemission spectroscopy (ARPES)~\cite{arpes}.  Several authors have
demonstrated that in this situation, an attractive exchange force in
the d-wave superconducting state binds an electron and a hole into a
pair with total spin 1~\cite{exciton,bl,artem1,mike:preprint}. This
bound state is seen as a resonance in a spin response.  The gross
features of the neutron resonance, such as a non-monotonic variation
of $\Omega_{\bf Q}$ with doping and persistence of the resonance in
the pseudogap phase can be understood if the resonance peak is viewed
as a collective mode\cite{bl,artem1,mike:preprint}.

Formally, a collective mode has the dispersion given by
Eq. (\ref{eq:massive-magnon}).  Its frequency is proportional to the
static part of the inverse spin susceptibility $(\xi^{-2} + |{\bf
q-Q}|^2)^{1/2}$~\cite{artem1}.  However, this time the velocity $c$ is
not constant but depends rather strongly on the wave vector ${\bf q}$.
This dependence can be understood by noting that the bound state
necessarily resides below the two-particle continuum
[Fig.~\ref{fig0}(c)].  For ${\bf q}$ along a zone diagonal, the
continuum starts at $E_{\rm min}({\bf q}) = |\Delta({\bf
k})|+|\Delta({\bf k+q})|$, where both momenta ${\bf k}$ and ${\bf
k+q}$ are at the Fermi surface, Fig.~\ref{fig0}(b).  As ${\bf q}$
moves away from $(\pi,\pi)$, ${\bf k}$ and ${\bf k+q}$ shift along the
Fermi surface towards the nodal points and the bound state inside the
gap is pushed to lower energies.  For ${\bf q=q}_0$ connecting the
nodal points, the two-particle spectrum is gapless, $E_{\rm min}({\bf
q}_0)=0$, so that the energy of the resonance must vanish (along with
its strength).  Obviously then, $\Omega_{\bf q}$ should decrease away
from $(\pi,\pi)$, at least near ${\bf q}_0$.

The above qualitative argument is quite robust and is based on two
premises only: (i) the Fermi surface contains hot spots (i.e., points
connected by the AF wave vector {\bf Q}) and (ii) the superconducting
gap has the $d_{x^2-y^2}$ symmetry.  A more quantitative description
of the resonance requires further assumptions.  Most calculations of a
spin response in a d-wave superconductor start with a model of free
fermions on a square lattice, add an exchange interaction and employ
the random-phase approximation (RPA) to compute spin susceptibility
\cite{exciton,bl,mike:preprint}.  The downward dispersion of the
resonance can then be obtained under favorable circumstances
\cite{mike:preprint}.  Unfortunately, such calculations must be done
numerically providing somewhat limited insight.  Here we present an
alternative, analytical approach to the problem.  It is based on the
observation that the behavior of a collective mode is determined
largely by low-energy fermion degrees of freedom and is therefore
insensitive to the physics at high energies.

Our point of departure is a macroscopic spin-fermion model that
describes low-energy fermionic quasiparticles (with a Fermi surface
inferred from ARPES) interacting with collective spin fluctuations.
The model is described by the effective action~\cite{acs}
\begin{eqnarray}
S &=&-\int_0^\beta d\tau \int_0^\beta d\tau' \sum_{{\bf k},\sigma}
c^\dagger_{{\bf k}\sigma} (\tau)
G^{-1}_0 ( {\bf k}, (\tau-\tau')) c_{{\bf k}\sigma}(\tau')       \nonumber \\
&& +  \frac{1}{2} \int_0^\beta d\tau \int_0^\beta d\tau'  
\sum_{{\bf q}} \chi_0^{-1} ({\bf q}, \tau) \, {\bf S}_{\bf q}(\tau) 
\cdot {\bf S}_{-{\bf q}} (\tau')\,                               \nonumber \\
&&  + g \int_0^\beta d\tau \sum_{{\bf q}}   \, 
{\bf s}_{\bf q}(\tau) \cdot {\bf S}_{-{\bf q}} (\tau)\, .        \label{sfm}
\end{eqnarray}
where $G^{-1}_0 ( {\bf k}, \tau) = \frac{\partial}{\partial \tau} -
{\bf v}_{\bf k}\!\cdot({\bf k}-{\bf k}_F)$ is the bare fermionic
propagator, and $ \chi_0^{-1} ({\bf q}, \tau)$ is the bare spin
susceptibility.  This model is viewed as a low-energy version of a
lattice, Hubbard-type model, obtained by integrating out high-energy
degrees of freedom.  Accordingly, it has a cut-off $\Lambda < E_F$ and
the bare $\chi_0 ({\bf q}, \tau)$ is determined by fermions with
energies exceeding $\Lambda$.  If, as we assume, nothing special
happens at high frequencies, $\chi_0 ({\bf q}, \Omega)$, the Fourier
transform of $\chi_0 ({\bf q}, \tau)$ will have an Ornstein-Zernike
form: $\chi_0 ({\bf q},\Omega ) =\chi _{0}/(\xi^{-2}+({\bf q}-{\bf
Q})^{2} -\Omega^2/v^2_s)$.

Previous studies of Eq. (\ref{sfm}) focused on renormalization of the
fermionic dispersion by spin-fermion interaction, and on the form of
the full $\chi ({\bf Q}, \Omega)$~\cite{artem1}.  Here we consider the
dynamical susceptibility at ${\bf q \neq Q}$.  To avoid unnecessary
complications, we restrict calculations to momenta along the zone
diagonal, ${\bf q} = (q,q)$.  We also neglect strong coupling effects,
assuming for simplicity that superconductivity is described by a
d-wave version of the BCS theory.  We have verified that
strong-coupling effects (which modify fermionic propagator at low
energies) result in quantitative, but not qualitative changes.

The full dynamical susceptibility $\chi ({\bf q}, \Omega)$ differs
from $\chi_0 ({\bf q}, \Omega)$ due to a bosonic self-energy $\Pi
({\bf q},\Omega)$:
\begin{equation}
\chi ({\bf q},\Omega ) =\frac{\chi _{0}}{\xi^{-2}+({\bf q}-{\bf Q})^{2} 
-\Omega^2/v^2_s -\Pi ({\bf q},\Omega)}.  
\label{chif}
\end{equation}
The static part of $\Pi ({\bf q},\Omega)$ (the contribution of the
low-energy fermions to the inverse correlation length $1/\xi$) is
small and, in fact, vanishes for linearized fermion dispersion.  On
the other hand, the frequency-dependent part of $\Pi ({\bf q},\Omega)$ 
is substantially nonzero in the normal state.  This is related to the 
fact that, for a Fermi surface with hot spots (as in YBCO and Bi2212)
a low-frequency spin excitation with momentum ${\bf q \approx Q}$ can 
decay into two fermions at the Fermi surface (Fig.~\ref{fig0}).  
This gives rise to a universal relaxational term in 
$\Pi ({\bf q},\Omega)$: 
\begin{equation}
\Pi ({\bf q},\Omega) \approx i\,{\rm Im}\Pi ({\bf q},\Omega)
= i |\Omega| \gamma_{\bf q}.
\label{pi}
\end{equation}
At small frequencies, $\Pi ({\bf q},\Omega) \propto |\Omega|$ is much
larger than the bare term ${\cal O}(\Omega^2)$ in the susceptibility.
It thus fully determines spin dynamics at low energies.  The prefactor
$\gamma_{\bf q}$ is obtained by calculating the imaginary part of the
particle-hole bubble.  In two dimensions, $\gamma_{\bf q} = 2\bar{g}/(\pi
v_x v_y)$, where ${\bar g} = g^2 \chi_0$ is the effective spin-fermion
coupling, and ${\bf v}=(v_x,v_y)$ is a ({\bf q}-dependent) Fermi
velocity at the hot spot ${\bf k_{\rm hs} (q)}$, whose components are
defined as $\epsilon_{\bf k} = v_x {\tilde k}_x + v_y {\tilde k}_y$,
$\epsilon_{\bf k +q} = - v_x {\tilde k}_x + v_y {\tilde k}_y$, where
${\bf \tilde{k} = k - k_{\rm hs} (q)}$~\cite{artem1}.  Ordinarily, the
{\bf q} dependence is weak and can be neglected.  However, for the
wave vector ${\bf q}_0$ connecting the nodes, the velocities at ${\bf
k}_{\rm hs}$ and ${\bf k}_{\rm hs} +{\bf q}_0$ are antiparallel,
therefore $v_y =0$ and $\gamma_{\bf q}$ diverges.  We have verified that
$\Pi({\bf q}_0,\Omega)$ is of order $\sqrt{\Omega}$.

Eqs. (\ref{chif}) and (\ref{pi}) imply that in the normal state 
\begin{equation}
{\rm Im}\chi ({\bf q},\Omega ) \propto \frac{x}{(1 + 
{\tilde q}^2 \xi^{2})^2 + x^2}
\label{4}
\end{equation}
where $x = \Omega/\gamma \xi^2$, and ${\tilde {\bf q}} = {\bf q} -
{\bf Q}$.  We see that spin response is (a) incoherent: no sharp
peak in $\chi^{\prime \prime} ({\bf q},\Omega)$ as a function of
frequency $\Omega$; (b) commensurate: $\chi^{\prime \prime} ({\bf
q}, \Omega)$ is peaked at ${\bf q = Q}$ at a fixed $\Omega$.  Both
results are in agreement with the data~\cite{n-st-neutrons}.

In a d-wave superconducting state, the bosonic self-energy is modified 
thanks to the opening of a superconducting gap.  It is now given by 
\begin{eqnarray}
\Pi ({\bf q},\Omega ) = \frac{i\gamma_{\bf q}}{2} 
\int\! 
\left(1 - 
\frac{\omega (\omega + \Omega) - \Delta^2_{\bf q}}
{\sqrt{[\omega^2 - \Delta^2_{\bf q}] [(\omega+\Omega)^2 - \Delta^2_{\bf q}]}}
\right) d\omega.
\label{5}
\end{eqnarray}
Here $\Delta_{\bf q}$ is the fermion gap at one of the hot spots
connected by ${\bf q}$, i.e., $\Delta_{\bf q} = \Delta ({\bf k})$,
such that both ${\bf k}$ and ${\bf k}+{\bf q}$ are at the Fermi surface.
By virtue of the $d_{x^2 - y^2}$ symmetry, $\Delta ({\bf k}) = -\Delta
({\bf k} +{\bf q})$, a condition we used in deriving Eq. (\ref{5}).

In the presence of a superconducting gap with the d-wave symmetry,
${\rm Im} \Pi ({\bf q},\Omega )$ vanishes discontinuously for
$\Omega<2|\Delta_{\bf q}|$.  By virtue of the Kramers-Kronig relation, this
discontinuity generates a nonzero ${\rm Re} \Pi({\bf q},\Omega)$
that is quadratic in $\Omega$ at low frequencies:
\begin{equation} 
\Pi ({\bf q},\Omega) 
\sim \gamma_{\bf q} \Omega^2/\Delta_{\bf q}
\hskip 3mm {\rm as}\ \ \Omega\to 0.
\label{6}
\end{equation}
It is also essential that, for any ${\bf q}$, $\Pi ({\bf q}, 0)=0$:
the opening of the d-wave gap does not change the magnetic correlation
length.  This result is not entirely surprising: spin-mediated d-wave
pairing involves only fermions from opposite sublattices and thus does
not affect the correlation of spin within the same sublattice.

Substitution of Eq. (\ref{6}) into (\ref{chif}) yields 
\begin{equation}
\chi ({\bf q},\Omega )\propto 
\frac{c^2_{\bf q}}{c^{2}_{\bf q} ( \xi^{-2} + {\tilde q}^{2})-\Omega ^{2}},
\hskip 5mm 
c_{\bf q}^2 = 
\frac{\Delta_{\bf q}}{\gamma_{\bf q}\xi^2}.
\label{chil}
\end{equation}
We see that in a $d-$wave superconductor the low energy spin
excitations are propagating, gapped magnon-like modes with the
dispersion
\begin{equation}
\Omega_{\bf q}^2 = c_{\bf q}^2 (\xi^{-2} + |{\bf q}-{\bf Q}|^{2})
\label{9}
\end{equation}
Eqs. (\ref{chil})--(\ref{9}) are meaningful only if $c_{\bf q}
\xi^{-1} < |\Delta_{\bf q}|$.  Otherwise the use of a quadratic form
for $\Pi({\bf q},\Omega)$ is not justified.  Strictly speaking, near
${\bf q=Q}$, $c_{\bf q} \xi^{-1} < \Delta_{\bf q}$ only at
sufficiently strong coupling, when fermion self-energy cannot be
neglected.  On the other hand, a quadratic frequency dependence of
${\rm Re}\Pi({\bf q},\Omega )$ at low $\Omega$ is merely a consequence of a
vanishing ${\rm Im} \Pi({\bf q},\Omega)$ for $\Omega < 2|\Delta_{\bf
q}|$, where at strong coupling $\Delta_{\bf q}$ should be understood
as the measured gap (i.e., a frequency where the spectral function has
a $\delta-$functional peak).  We have explicitly verified that
inclusion of strong coupling corrections into (\ref{5}) only changes
the overall factor in (\ref{6}).

\begin{figure}[tbp]
\centerline{\epsfxsize=0.47\columnwidth 
\epsffile{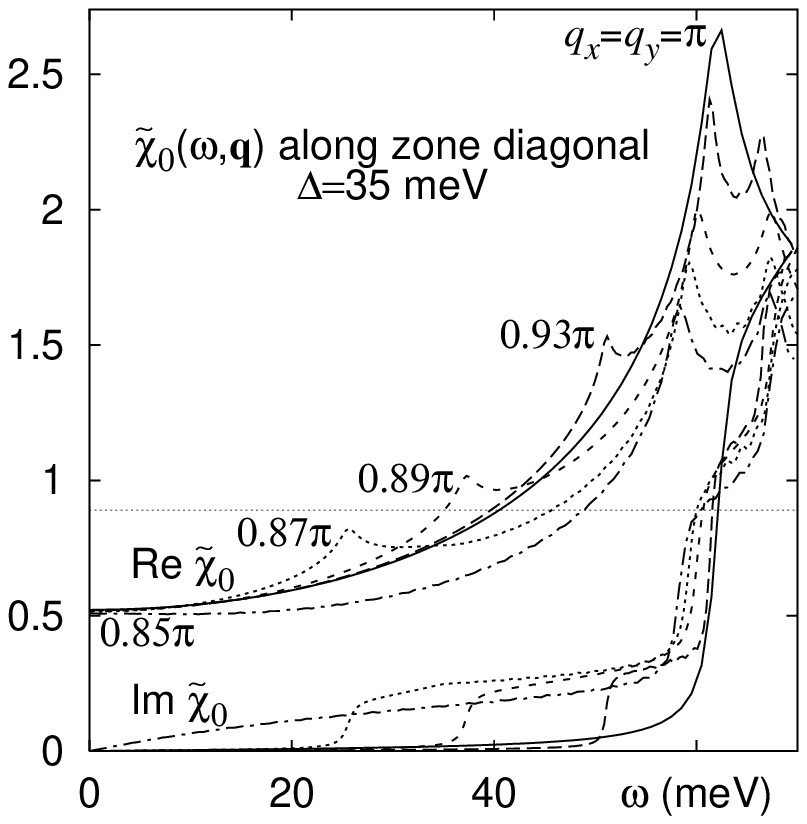}
\epsfxsize=0.47\columnwidth 
\epsffile{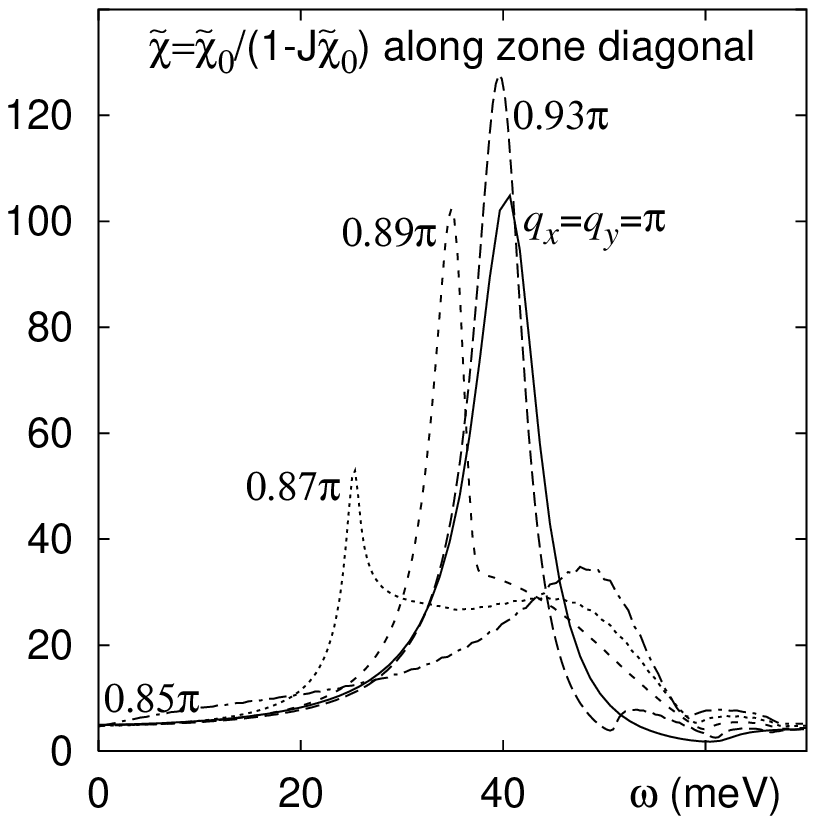}}
\caption{Left: real and imaginary parts of the numerically evaluated
particle-hole bubble in a $d-$wave superconductor,
${\tilde\chi}_0({\bf q},\omega)$, at various ${\bf q} = (q,q)$, using
a tight-binding fermion dispersion inferred from ARPES. The
low-frequency features of ${\tilde\chi}_0({\bf q},\omega)$ are in
agreement with our analytical results.  Right: imaginary part of the
corresponding full dynamical susceptibility, ${\rm Im}\chi(\Omega,{\bf
q})$.  Observe that the resonance moves to lower frequencies when
${\bf q}\neq{\bf Q}=(\pi,\pi)$.}
\label{fig1}
\end{figure}

To verify that our analytical approach [based on a linearized fermion
dispersion around ${\bf k}_{\rm hs} ({\bf q})$] captures the
essential features of the spin susceptibility, we present in
Fig.~\ref{fig1} numerical results for the bare particle-hole
susceptibility ${\tilde \chi}_0({\bf q},\Omega)$ and its RPA
counterpart ${\tilde \chi} ({\bf q},\Omega)$ ${\tilde \chi}({\bf
q},\Omega) = {\tilde \chi}_0({\bf q},\Omega)/[1 - J {\tilde \chi}_0
({\bf q},\Omega)]$.  Both are calculated for a $d_{x^2-y^2}$ BCS
superconductor with a tight-binding dispersion $\epsilon_k$ inferred
from ARPES data\cite{mike:preprint}.  We expect that, at low enough
$\Omega$, $\Pi ({\bf q},\Omega)$ matches ${\tilde \chi}_0({\bf
q},\Omega)$, modulo a regular function of $q$.  Similarly, $\chi ({\bf
q},\Omega)$ should agree, at low frequencies, with ${\tilde \chi}
({\bf q}, \Omega)$.

In Fig.~\ref{fig1}, we see that ${\rm Im} {\tilde \chi}_0({\bf q},
\Omega)=0$ for $\Omega<2|\Delta_{\bf q}|$ and has a finite jump at
$2\Delta_{\bf q}$, while ${\rm Re}{\tilde \chi}_0({\bf q},\Omega)$ is
quadratic in $\Omega$ at low frequencies, in agreement with what we
obtained for $\Pi ({\bf q}, \Omega)$.  The peaks in ${\rm Re}{\tilde
\chi}_0({\bf q},\Omega)$ at $2\Delta_{\bf q}$ are logarithmic
singularities associated with the discontinuity of ${\rm Im} {\tilde
\chi}_0(2\Delta_{\bf q})$.  They are softened because we used complex
frequencies with a small imaginary part.  These singularities are
indeed present in $\Pi ({\bf q}, \Omega)$.  A strong peak in the RPA
susceptibility ${\tilde \chi}({\bf q},\Omega)$ is the resonance
located at $1=J {\rm Re}{\tilde \chi}_0({\bf q},\Omega)$.  This
expression corresponds to $\Pi ({\bf q},\Omega) = \xi^{-2} + ({\bf
q}-{\bf Q})^2$ in our analytical approach.  We have also verified that
the static susceptibility ${\tilde \chi}_0({\bf q},0)$ does not change
between normal and superconducting states (i.e., $\xi$ is not
renormalized).  High-frequency features in Fig~\ref{fig1}---not not
captured by our long-wavelength approximation---are lattice effects
sensitive to the details of the fermion dispersion $\epsilon_{\bf k}$
far from the Fermi surface.

We now analyze Eq. (\ref{9}). If $\Delta_{\bf q}$ and $\gamma_{\bf q}$
were independent of momentum, $c_{\bf q}$ would be constant and
$\Omega_{\bf q}$ would be a converntional magnon-like dispersion
(\ref{eq:massive-magnon}). In this situation, the spin resonance would
remain commensurate and exist only at $\Omega > \Omega_{\bf Q} =
c_{\bf Q}/\xi$.  However, in a d-wave superconductor, $\Delta_{\bf q}$
decreases and $\gamma_{\bf q}$ increases when ${\bf q}$ deviates from
${\bf Q}$.  This effect accounts for the downward shift of
$\Omega_{\bf q}$.  Furthermore, as $c_{\bf q}$ vanishes at ${\bf q\to
q}_0$, the resonance frequency $\Omega_{\bf q}\to 0$ regardless of the
spin correlation length $\xi$.  Quite generally, near ${\bf q}_0$,
${\rm Im} \chi({\bf q}, \Omega)$ should have a peak at a frequency
$\Omega_{\bf q} < \Omega_{\bf Q}$.

The behavior near ${\bf q=Q}$ is more complicated and depends on the
values of $\xi$ and ${\bf q}_0$.  We can roughly approximate the
momentum dependence of $c_{\bf q}$ as 
\begin{equation}
c_{\bf q}^2 \approx 1 - ({\tilde q}/{\tilde q}_0)^2,
\hskip 5mm 
{\tilde q} = |{\bf q} - {\bf Q}|.
\label{eq:cq}
\end{equation} 
While the specific functional form we adopt here for $c_{\bf q}$ is
unimportant for capturing the qualitative behavior of the neutron
resonance dispersion, the form adopted in Eq. (\ref{eq:cq}) is
appropriate for a magnetically mediated d-wave superconductivity in
which $\Delta_{\bf q}$ is largest at ${\bf q=Q}$~\cite{acs}.

Upon substituting this form into Eq. (\ref{9}), we find
\begin{equation}
\frac{\Omega_{\bf q}}{\Omega_{\bf Q}}  
= 1 - \left(\frac{\tilde q}{{\tilde q}_0}\right)^2 (1 - ({\tilde q}_0 \xi)^2) 
 - \left(\frac{\tilde q}{{\tilde q}_0}\right)^4 ({\tilde q}_0 \xi)^2
\label{10}  
\end{equation}
When ${\tilde q}_0 \xi$ is small, the dispersion is negative for all
momenta. Also, the residue of the peak in Eq. (\ref{chil}) scales as
$c_{\bf q}$, i.e., it decreases and eventually vanishes as ${\bf q}$
approaches ${\bf q}_0$.

\begin{figure}[tbp]
\centerline{
\epsfxsize=0.47\columnwidth
\epsffile{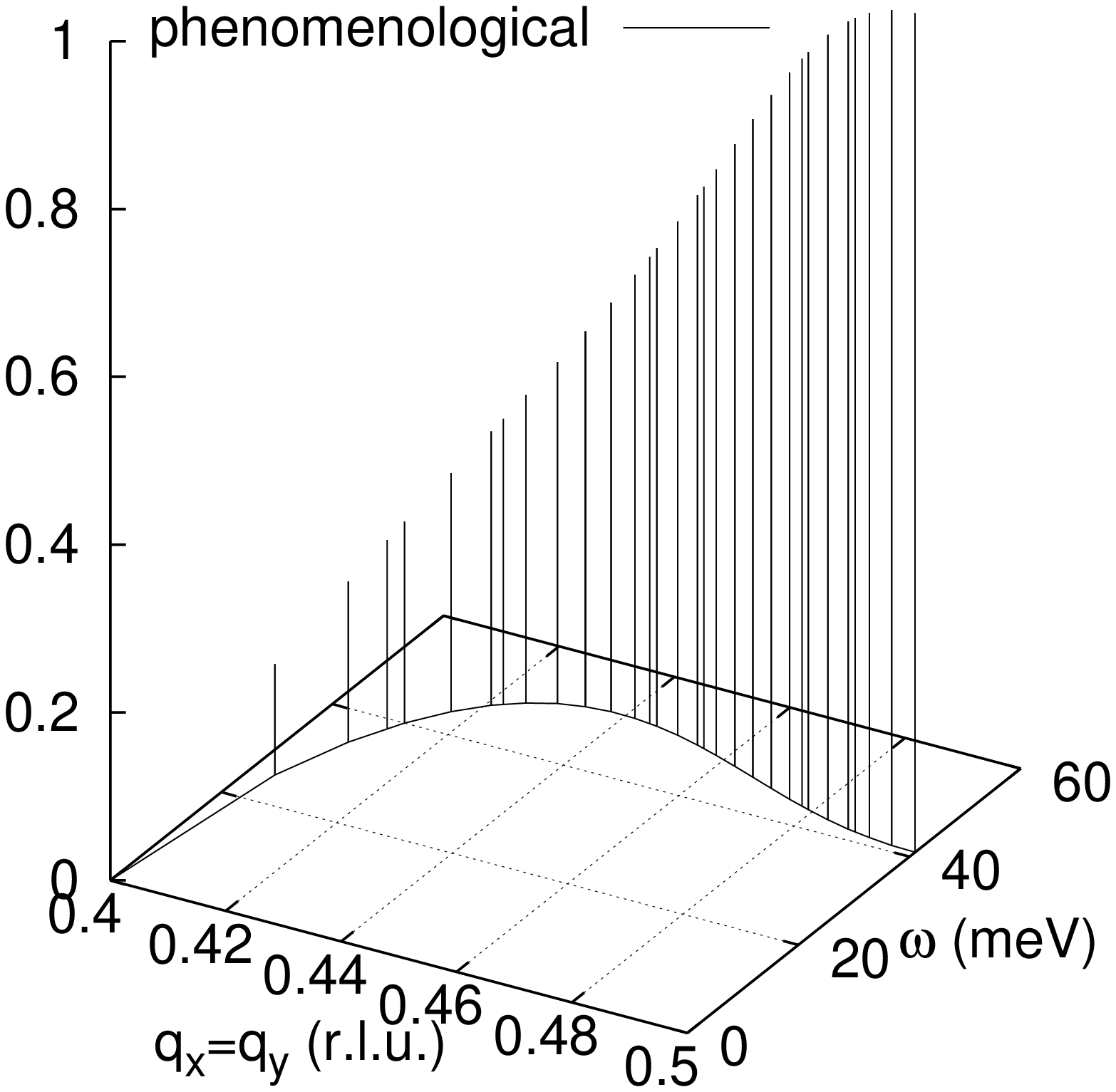}
\epsfxsize=0.47\columnwidth
\epsffile{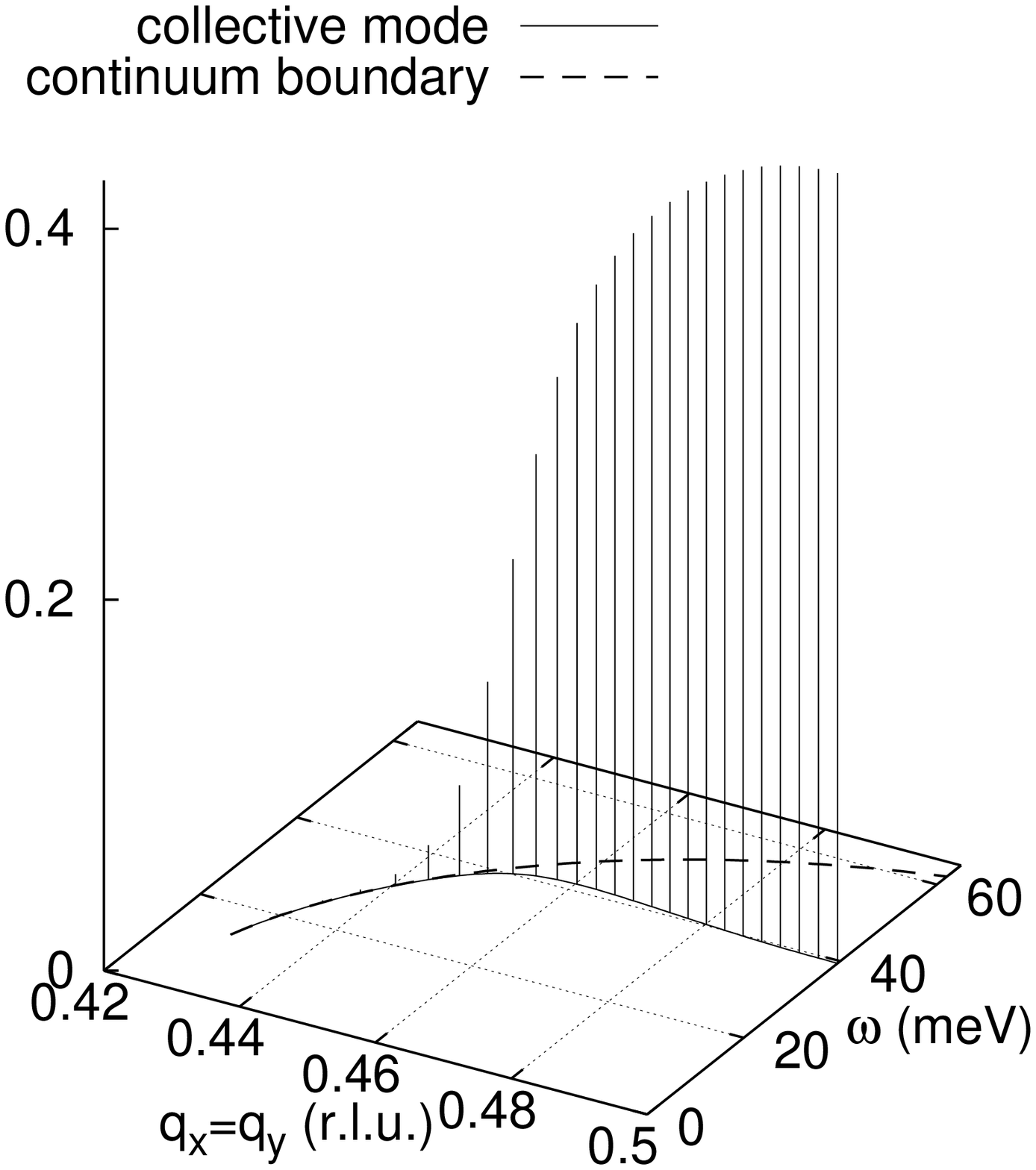}}
\caption{Location and intensity of the resonance peak obtained from
Fig.~\protect\ref{fig1} (left figure) and from
Eqs. (\protect\ref{chil}) and (\protect\ref{10}) (right figure).}
\label{figdisp}
\end{figure}

In Fig.~\ref{figdisp} we plot $\Omega_{\bf q}$ (\ref{10}) and the
intensity of the peak for ${\tilde q}_0 \xi =2$. This value is
consistent with optimally doped Bi2212 where $|{\bf q}_0| \approx 0.3
\pi/a$~\cite{peter}) while $\xi = a - 2a$~\cite{david}.
$\Omega_{\bf q}$ in (\ref{10}) is rather flat near ${\bf Q}$ and
rapidly drops away from ${\bf Q}$.  The residue of the peak at the
downturn is already much smaller than its value at ${\bf Q}$.  Both
these features are consistent with the data \cite{Keimer3}.

For comparison, we also present in Fig.~\ref{figdisp} the dispersion
and intensity of the collective mode obtained in the RPA calculation,
Fig.~\ref{fig1}.  It exhibits qualitatively similar behavior which we
interpret as evidence that the dispersion of the peak is insensitive
to the details of the fermion dispersion at high energies.  Note also
that our estimate of the wave vector at which the energy and intensity
of the neutron peak vanishes, ${\bf q}_0 \approx (0.8\pi,0.8\pi)$, is
roughly consistent with the corresponding value found in neutron
scattering, ${\bf q} = (\pi, 0.8\pi)$.  Recall that, to simplify the
analysis, we have only considered momenta along zone diagonal.

Finally, we comment on the limit ${\tilde q}_0 \xi\to\infty$ relevant
to the underdoped side.  In this limit, $\Omega_{\bf q}$ first
increases with ${\tilde q}$ and drops only very near ${\bf q = q_0}$.
This implies that the resonance continuously evolves into the magnon
of the AF state.  It would be of interest to verify experimentally
whether such evolution takes place.

To summarize, we have demonstrated that the experimentally observed
downturn of the resonant frequency \cite{Keimer3} away from
$(\pi,\pi)$, accompanied by a rapid decrease of the peak intensity,
occurs rather naturally when the resonance peak is interpreted as a
collective spin excitation in a d-wave superconductor.  The unusual
dispersion of the peak is related to a variation of the
superconducting gap along the Fermi surface.  We have argued that the
resonant frequency vanishes at a certain ${\bf q=q}_0$ that connects
nodal points at the Fermi surface.  Close to the commensurate wave
vector ${\bf Q=(\pi,\pi)}$, the dispersion depends on the spin
correlation length: If the latter is small, $|{\bf q}_0-{\bf Q}|\xi
\ll 1$, the resonance frequency $\Omega_{\bf q}$ decreases with $|{\bf
q-Q}|$.  Conversely, when $|{\bf q}_0-{\bf Q}|\xi \gg 1$, $\Omega_{\bf
q}$ increases away from {\bf Q}, reminiscent of a magnon dispersion in
a disordered antiferromagnet.

We acknowledge useful conversations with Ar. Abanov, J.C. Campuzano,
P. Dai, M.R. Norman and S. Sachdev.  The research was supported in
part by the NSF Grant No.~DMR-9979749, the US DOE Grant
No. W-31-109-ENG-38 and DE-FG02-90ER40542.  A.Ch. and O.T. thank the
Materials Science Division of Argonne National Laboratory for
hospitality during their stay at Argonne.


\end{document}